\documentclass[review]{elsarticle}

\newcommand{\F}{\mathbb{F}}

\usepackage{mathrsfs}
\usepackage{amssymb}
\usepackage{amsmath}
\usepackage{amsthm}
\usepackage{latexsym}
\usepackage{indentfirst}
\usepackage{enumitem}
\usepackage{anysize}
\usepackage{bbm}

\usepackage{hyperref}

\journal{Journal of \LaTeX\ Templates}
\begin{document}

\begin{frontmatter}

\title{ Three-weight codes and the quintic construction}



\author{Yan Liu}
\address{School of Mathematical Sciences, Anhui University, Hefei, Anhui Province 230601, P.R. China}
\fntext[myfootnote]{The author is supported by NNSF of China (61672036),
Technology Foundation for Selected Overseas Chinese Scholar, Ministry of Personnel of China (05015133), the Open Research Fund of National Mobile Communications Research Laboratory, Southeast University (2015D11) and Key projects of support program for outstanding young talents in Colleges and Universities (gxyqZD2016008).}
\ead{liuyan2612@126.com}

\author[mymainaddress]{Minjia Shi$^{*}$\fnref{myfootnote}}
\ead{smjwcl.good@163.com}

\author[mysecondaryaddress]{Patrick Sol\'e}
\cortext[mycorrespondingauthor]{Corresponding author}
\ead{sole@enst.fr}

\address[mymainaddress]{Key Laboratory of Intelligent Computing $\&$ Signal Processing,
Ministry of Education, Anhui University No. 3 Feixi Road, Hefei Anhui Province 230039, P. R. China, National Mobile Communications Research Laboratory, Southeast University and School of Mathematical Sciences of Anhui University,
Anhui, 230601, P. R. China}
\address[mysecondaryaddress]{CNRS/LAGA, University Paris 8, 93 526 Saint-Denis, France}

\begin{abstract}
We construct a class of three-Lee-weight and  two infinite families of five-Lee-weight codes over the ring $R=\mathbb{F}_2+v\mathbb{F}_2+v^2\F_2+v^3\F_2+v^4\F_2,$ where
$v^5=1.$ The same ring occurs in the quintic construction of binary quasi-cyclic codes.
They have the algebraic structure of abelian codes. Their Lee weight distribution is computed by using character sums.
 Given a linear Gray map, we obtain three families of binary abelian codes with few weights. In particular, we obtain a class of three-weight codes which are optimal.
 Finally, an application to secret sharing schemes is given.
\end{abstract}

\end{frontmatter}


\section{Introduction}
Let $\F_p$ denote the finite field with $p$ elements. An $[n,k,d]$ linear code $C$ over $\F_p$ is an $k$-dimensional subspace of $\F_p^n$ with minimum Hamming distance $d.$ An $[n,k,d]$ linear code is called optimal if no $[n,k,d+1]$ code exists.
A classical construction of codes over finite fields called {\bf trace codes} is documented in \cite{D,DLLZ,SLS1,SWLP}. Many known codes \cite{DY, HY1,ZD1,ZD2,ZDL}
can be produced by this construction.


In a series of papers \cite{SLS1,SLS2,SLS3,SWLP}, we have extended the notion of trace codes from fields to rings as follows. If $R$ is a finite ring, and $R_m$ an extension of $R$ of degree $m,$ $R_m^*$ denotes the group of units of $R_m$, we construct
a trace code with a defining set $\mathcal{L}=\{d_1,d_2,\dots,d_{n'}\}\subseteq R_m^*$ by the formula
$$C_\mathcal{L}=\{(Tr_m(xd_1),Tr_m(xd_2),\dots,Tr_m(xd_{n'})):x\in R_m \}=\{(Tr_m(xd))_{d \in \mathcal{L}}: x\in R_m\},$$ where $Tr_m()$ is a linear function from $R_m$ down to $R.$
By varying $\mathcal{L}$ and $R,$ various codes can be constructed. We can summarize this research program as shown below.

\begin{itemize}
\item[\cite{SLS1}] $\mathcal{L}=R_m^*,$ $R=\F_2+u\F_2$;
\item[\cite{SLS2}] $L=R_m^*,$ $R=\F_2+u\F_2+v\F_2+uv\F_2$;
\item[\cite{SWLP}] $[R_m^*:\mathcal{L}]=2,$ $R=\F_p+u\F_p$;
\item[\cite{SLS3}] $\mathcal{L}=D+u \F_{p^m},$ $[R_m^*:\mathcal{L}]=(p-1)gcd(N,\frac{p^m-1}{p-1}),$ $R=\F_p+u\F_p$.
\end{itemize}
Note that all rings in the above list are chain rings, except the second one.
In the present paper, we will consider another semi-local ring $\F_2+v\F_2+v^2\F_2+v^3\F_2+v^4\F_2$, where $v^5=1.$ This ring occurs in the quintic construction of binary quasi-cyclic codes
\cite{BNS}. While the binary codes we describe are not obtained by that construction, they are still quasi-cyclic of co-index $5.$
By the generic method of our research program, we obtain a family of optimal binary codes by using a linear Gray map.
 Furthermore,
an application to secret sharing schemes is sketched out.

The rest of this paper is organized as follows. The next section describes the basic notations and defines a Gray map, which will be needed in Section 4. Section 3 shows that the codes are abelian. Section 4 gives the main results in this paper, the Lee weight distribution of our codes. Furthermore, we show that the Gray images of three-Lee-weight codes are optimal. Section 5 determines the minimum Lee distance of their dual codes. The codes we constructed have applications in secret sharing schemes in section 6. We will sum up all we have done throughout this paper in section 7, and make some conjectures for future research.

\section{Preliminaries}
\subsection{Rings}

Throughout this paper, we let $\F_2$ be a finite field with two elements, i.e., $\F_2=\{0,1\}$. Denote by $R$ the commutative ring $\F_2+v\F_2+v^2\F_2+v^3\F_2+v^4\F_2$,
constructed via $v^5=1$. $R$ is a ring of size $2^5$ with characteristic 2. Because the factorization of $v^5-1$ into irreducible factors is $(v-1)(1+v+v^2+v^3+v^4),$ the ring $R$ has two
maximal ideals, namely, $(1+v)=\{(1+v)(a_0+a_1v+a_2v^2+a_3v^3+a_4v^4):a_i\in \F_2,i=0,1,2,3,4\}$ and $(1+v+v^2+v^3+v^4)=\{0, 1+v+v^2+v^3+v^4 \}$. Thus it is a non-local, non-chain principal ideal ring.
Given a positive integer $m$, we can construct the ring extension $R_m=\F_{2^m}+v\F_{2^m}+v^2\F_{2^m}+v^3\F_{2^m}+v^4\F_{2^m}.$ Let $\epsilon\in \F_{16}$ be a root of the irreducible polynomial $1+v+v^2+v^3+v^4$ in $\F_2$, then by a simple calculation, we get the
factorization of $v^5-1$ as follow:
\begin{eqnarray*}
  v^5-1 &=&(1+v)(1+v+v^2+v^3+v^4)~~~~~~~~~~~ \mathrm{in}~ \F_2,\\
   &=& (1+v)(1+\omega^2v+v^2) (1+\omega v+v^2)~~~  \mathrm{in}~ \F_4=\{0,1,\omega,\omega^2\},\\
  &=& \prod_{i=0}^4(v+\epsilon^i) ~~~~~~~~~~~~~~~~~~~~~~~~~~~~~~~~\mathrm{in}~ \F_{16}=\{\nu_0+\nu_1\epsilon+\nu_2\epsilon^2+\nu_3\epsilon^3:\nu_i\in \F_2,i=0,1,2,3\}.
\end{eqnarray*}
Hence, by using Chinese Remainder Theorem, the ring $R_m$ is seen to be isomorphic to
$\F_{2^m}\bigoplus \F_{16^m} $ when $m$ is odd, and $\F_{2^m}\bigoplus \F_{4^m}\bigoplus \F_{4^m} $ when $m$ is singly-even, and
$\F_{2^m}\bigoplus\F_{2^m}\bigoplus\F_{2^m}\bigoplus\F_{2^m}\bigoplus\F_{2^m}$ when $m$ is doubly-even.
Here $R_m^*$ denotes the group of units in $R_m$, and $\F_{2^m}^*$ denotes the multiplicative cyclic group of nonzero elements of $\F_{2^m}.$ Likewise, we have $R_m^*\cong\F_{2^m}^*\bigoplus \F_{16^m}^* $ when $m$ is odd, and $R_m^*\cong\F_{2^m}^*\bigoplus \F_{4^m}^*\bigoplus \F_{4^m}^* $ when $m$ is singly-even,
and $R_m^*\cong\F_{2^m}^*\bigoplus \F_{2^m}^*\bigoplus\F_{2^m}^*\bigoplus\F_{2^m}^*\bigoplus\F_{2^m}^*$ when $m$ is doubly-even.

 Let $tr_m()$ be the trace function from $\F_{2^m}$ to $\F_2$, namely, for any $\ell\in \F_{2^m}, tr_m(\ell)=\ell+\ell^2+\dots+\ell^{2^{m-1}}$. Similar to the definition of $tr_m()$, we define the \emph{Trace function,} denoted by $Tr_m()$, over $R_m.$ \vspace*{0.3cm}

\noindent{\bf  Definition 2.1} For any $a=a_0+a_1v+a_2v^2+a_3v^3+a_4v^4\in R_m,$ where $a_i\in \F_{2^m}, i=0,1,2,3,4,$ the \emph{trace} $Tr_m(a)$ of $a$ over $R$ is defined by
$$Tr_m(a_0+a_1v+a_2v^2+a_3v^3+a_4v^4)=tr_m(a_0)+tr_m(a_1)v+tr_m(a_2)v^2+tr_m(a_3)v^3+tr_m(a_4)v^4.$$

It is well known that the trace function $tr_m()$ is a linear transformation from $\F_{2^m}$ onto $\F_2$. So it is immediate to check that $R$-linearity of $Tr_m()$ follows from the $\F_2$-linearity of $tr_m()$.


\subsection{Codes and Gray map}
  A {\bf linear code} $C$ over $R$ of length $n$ is an $R$-submodule of $R^n$. The elements of a such code are called its {\bf codewords}. For $x=(x_1,x_2,\dots,x_n),y=(y_1,y_2,\dots,y_n)\in R^n$, their standard inner product
  is defined by $\langle x,y\rangle=\sum_{i=1}^nx_iy_i$, where the operation is performed in $R$.
 The {\bf dual code} $C^\perp$ of a linear code $C$, over $R$, consists of all vectors of $R^n$
which are orthogonal to every codeword in $C$, that is, $C^\perp=\{y\in R^n|\langle x,y\rangle =0, \forall x\in C\}.$

We define the following linear Gray map which takes a linear code over $R$ of length $n$ to a binary linear code of length $5n$.\vspace*{0.3cm}

\noindent{\bf  Definition 2.2}
The Gray map $\Phi$ from $R$ to $\F_2^5$ is defined as
$$\Phi(a_0+a_1v+a_2v^2+a_3v^3+a_4v^4)=(a_0,a_1,a_2,a_3,a_4),$$
where $a_i\in \F_2,i=0,1,2,3,4.$

This map $\Phi$ can be extended to $R^n$ in an obvious way. Define the Lee weight of an element $a_0+a_1v+a_2v^2+a_3v^3+a_4v^4$ of $R$ as $w_L(a_0+a_1v+a_2v^2+a_3v^3+a_4v^4)=w_H(\Phi(a_0+a_1v+a_2v^2+a_3v^3+a_4v^4))=
w_H((a_0,a_1,a_2,a_3,a_4))=\sum_{i=0}^4w_H(a_i),$ where $w_H$ denotes the usual Hamming weight. Then $\Phi$ is a distance preserving isometry from $(R^n,d_{L'})$ to $(\F_2^{5n},d_H)$, where $d_{L'}$ and $d_H$ denote the Lee and Hamming distance in $R^n$ and $\F_2^{5n}$, respectively. What is more, if $C$ is a linear code over $R$ with parameters $(n,2^k,d)$, then $\Phi(C)$ is a linear code over $\F_2$ with parameters $[5n,k,d]$.
It is immediate that linear codes of length $n$ over $R$ are mapped into $n$-quasi-cyclic binary codes of length $5n.$

 Given a finite abelian group $G,$ a code over $R$ is said to be {\bf abelian} if it is an ideal of the group ring $R[G].$
 Recall that the ring $R[G]$ is defined on functions from $G$ to $R$ with pointwise addition as addition, and convolution product as multiplication.
  Concretely it is the set of all formal sums $f=\sum_{h \in G} f_h X^h,$ where $X$ an undeterminate, with addition and multiplication defined as follows.
  If $f,g$ are in $R[G],$  we write
  $$f+g=\sum_{g \in G} f_h+g_h X^h,$$ and

  $$ fg =\sum_{h \in G} (\sum_{r+s=h} f_r g_s)X^h.$$
 In other words, the coordinates of $C$ are indexed by elements of $G$, and $G$ acts regularly on this set.
 In the special case when $G$ is cyclic, the code is a cyclic code in the usual sense \cite{MS}.


\section{Symmetry}
First, for $a\in R_m$ define the vector $ev(a)$ by the evaluation map $ev(a)=(Tr_m(ax))_{x\in R_m^* }.$ Define the code $\mathcal{C}(m,2,L)$ of length $L$ by the formula $\mathcal{C}(m,2,L)=\{ev(a):a\in R_m\}$. Thus $\mathcal{C}(m,2,L)$ is a linear code over $R$ and $\Phi(\mathcal{C}(2,m,L))$ is a linear binary code of length $5L.$\vspace*{0.3cm}


\noindent{\bf Proposition 3.1} If $L=|R_m^*|$, then the group $R_m^*$ acts regularly on the coordinates of $\mathcal{C}(m,2,L).$

\begin{proof}
For any $w,v \in L$ the change of variables $ x\mapsto (v/w)x$ maps $w$ to $v.$ This transformation defines thus a transitive action of $L$ on itself.
Given an ordered pair $(w,v)$ this transformation is unique,
hence the action is regular.
\end{proof}
The code $\mathcal{C}(m,2,|R_m^*|)$ is thus an {\em abelian code} with respect to the group $R_m^*.$ In other words, it is an ideal of the group ring $R[R_m^*].$ As observed in the previous section $R_m^*$ is a not cyclic group, hence $\mathcal{C}(m,2,L)$ may be not cyclic.

\section{The Lee Weight of $\mathcal{C}(m,2,L)$}
For convenience, we let $s=5|R_m^*|$. If $y=(y_1,y_2,\dots,y_s)\in \mathbb{F}_2^s,$ let $$\theta(y)=\sum_{j=1}^s(-1)^{y_j}.$$ For simplicity, we let $\Theta(a)=\theta(\Phi(ev(a))).$
In order to determine the Lee weight of the codewords of $\mathcal{C}(m,2,L)$, we first recall the following two lemmas.\vspace*{0.3cm}

\noindent\textbf{Lemma 4.1}~\cite{SWLP}\label{5.1} For all $y=(y_1,y_2,\dots,y_s)\in \mathbb{F}_2^s,$ we have
$$2w_H(y)=s-\sum_{j=1}^{s}(-1)^{y_j}.$$

\noindent\textbf{Lemma 4.2}~\cite{MS} If $z \in \mathbb{F}_{2^m}^*,$ then $$\sum\limits_{x\in \mathbb{F}_{2^m}}(-1)^{tr_m(z x)}=0.$$\\

According to Lemma 4.1, for $ev(a)\in \mathcal{C}(m,2,L)$, by definition of the Gray map, we have
\begin{equation}\label{1}
  2w_L(ev(a))=2w_H(\Phi(ev(a)) ) = s-\Theta(a).
\end{equation}

\subsection{ The first family of codes $\mathcal{C}(m,2,L_1)$ when $m$ is odd }

In the previous section, we know $R_m^*\cong \F_2^*\bigoplus\F_{16^m}^*$ when $m$ is odd.
A simple calculation shows that $$R_m^*=\Bigg{\{}\sum_{i=0}^4x_iv^i: \sum_{i=0}^4x_i\neq 0, (x_1+x_2+x_3+x_4,x_0+x_1,x_3+x_4,x_0+x_1+x_2+x_3)\neq(0,0,0,0) \Bigg{\}}.$$
For convenience, we adopt the following notations unless otherwise stated
in this section. Set $I=\sum_{i=0}^4x_i,I_1=x_1+x_2+x_3+x_4,I_2=x_0+x_1,I_3=x_3+x_4,I_4=x_0+x_1+x_2+x_3$.
Then $R_m^*=\{\sum_{i=0}^4x_iv^i: I\neq 0, (I_1,I_2,I_3,I_4)\neq(0,0,0,0),I,I_1,I_2,I_3,I_4\in \F_{2^m} \}.$

 We are now ready to discuss the Lee weight of the
codewords of the abelian codes introduced above.

\noindent{\bf Theorem 4.3} Let $m$ be odd, then the set $\mathcal{C}(m,2,L_1)$ is a three-Lee-weight linear code of length $L_1=(2^m-1)(2^{4m}-1)$ and its weight distribution is given in Table I.
\begin{center}$\mathrm{Table~ I. }~~~\mathrm{weight~ distribution~ of}~ \mathcal{C}(m,2,L_1) $\\
\begin{tabular}{cccc||cc}
\hline
  Weight&&   & & Frequency  \\
  \hline

  0        & &   & & 1\\
  $5\times(2^{5m-1}-2^{4m-1}-2^{m-1})$        & &   &              &$(2^{m}-1)(2^{4m}-1)$\\
    $5\times(2^{5m-1}-2^{4m-1})$        & &   &              &$2^{4m}-1$\\
  $5\times(2^{5m-1}-2^{m-1})$  &    & &       &$2^m-1$ \\
  \hline
\end{tabular}
\end{center}

\begin{proof}
Set $x=x_0+x_1v+x_2v^2+x_3v^3+x_4v^4\in R_m^*$, then $I\neq0,(I_1,I_2,I_3,I_4)\neq(0,0,0,0)$.
For $a=a_0+a_1v+a_2v^2+a_3v^3+a_4v^4\in R_m$, we have
\begin{eqnarray*}
  ax &=&[(a_0+a_1+a_2+a_3+a_4)I+(a_0+a_4)I_1+
(a_3+a_4)I_2+(a_2+a_3)I_3+(a_1+a_2)I_4]\\
&& +[(a_0+a_1+a_2+a_3+a_4)I+(a_0+a_1)I_1+
(a_0+a_4)I_2+(a_3+a_4)I_3+(a_2+a_3)I_4]v \\
&&+ [(a_0+a_1+a_2+a_3+a_4)I+(a_1+a_2)I_1+
(a_0+a_1)I_2+(a_0+a_4)I_3+(a_3+a_4)I_4]v^2\\
&&+  [(a_0+a_1+a_2+a_3+a_4)I+(a_2+a_3)I_1+
(a_1+a_2)I_2+(a_0+a_1)I_3+(a_0+a_4)I_4]v^3 \\
&&+  [(a_0+a_1+a_2+a_3+a_4)I+(a_3+a_4)I_1+
(a_2+a_3)I_2+(a_1+a_2)I_3+(a_0+a_1)I_4]v^4\\
&=:& A_0+A_1v+A_2v^2+A_3v^3+A_4v^4.
\end{eqnarray*}
So $$\Phi(Tr_m(ax))=(tr_m(A_0), tr_m(A_1),tr_m(A_2), tr_m(A_3),tr_m(A_4)).$$
Taking character sums, yields
$$\Theta(a) =\sum_{i=0}^{4}\sum_{I\in \F_{2^m}^*} \sum_{ (I_1,I_2,I_3,I_4)\neq(0,0,0,0),I_1,I_2,I_3,I_4\in \F_{2^m}}(-1)^{tr_m(A_i)  }.$$

If $a=a_0+a_1v+a_2v^2+a_3v^3+a_4v^4\in R_m^*$, then $\sum_{i=0}^4a_i\neq 0$ and $ (a_1+a_2+a_3+a_4,a_0+a_1,a_3+a_4,a_0+a_1+a_2+a_3)\neq(0,0,0,0) $, and $ \Theta(a)=5.  $
By Equation (1), we get $w_L(ev(a))=5\times(2^{5m-1}-2^{4m-1}-2^{m-1})$.

Note that $\sum_{i=0}^4a_i= 0$ and $ (a_1+a_2+a_3+a_4,a_0+a_1,a_3+a_4,a_0+a_1+a_2+a_3)=(0,0,0,0)$, i.e., $a=0$, corresponds to the zero codewords, and then its Lee weight is 0.

If $\sum_{i=0}^4a_i= 0$ and $ (a_1+a_2+a_3+a_4,a_0+a_1,a_3+a_4,a_0+a_1+a_2+a_3)\neq(0,0,0,0)$, then $a=a_0+a_1v+a_2v^2+a_3v^3+a_4v^4$ is a zero-divisor, and  $ \Theta(a)=-5\times(2^m-1). $
By Equation (1), we get $w_L(ev(a))=5\times(2^{5m-1}-2^{4m-1})$.

If $a=a_0+a_1v+a_2v^2+a_3v^3+a_4v^4$ satisfies $\sum_{i=0}^4a_i\neq0$ and $ (a_1+a_2+a_3+a_4,a_0+a_1,a_3+a_4,a_0+a_1+a_2+a_3)= (0,0,0,0)$, then we can claim $a=\alpha(1+v+v^2+v^3+v^4)$, where $\alpha\in \F_{2^m}^*.$
By calculation, we know
\begin{eqnarray*}
  \Theta(a) &=&5\sum_{I\in \F_{2^m}^*}\sum_{(I_1,I_2,I_3,I_4)\neq(0,0,0,0),I_1,I_2,I_3,I_4\in \F_{2^m}}(-1)^{tr_m(\alpha I) }=-5\times(2^{4m}-1).
\end{eqnarray*}
Hence we get $w_L(ev(a))=5\times(2^{5m-1}-2^{m-1})$ by Equation (1).
\end{proof}\vspace*{0.2cm}

According to Theorem 4.3, we have constructed a binary linear code of length $s=5\times(2^m-1)(2^{4m}-1)$, of dimension $5m$, with three nonzero weights $w_1<w_2<w_3$ of values
$$w_1=5\times(2^{5m-1}-2^{4m-1}-2^{m-1}), ~~~w_2=5\times(2^{5m-1}-2^{4m-1}),~~~w_3=5\times(2^{5m-1}-2^{m-1}),$$
with respective frequencies
$$f_1=(2^{m}-1)(2^{4m}-1),~~~f_2=2^{4m}-1,~~~f_3=2^m-1.$$

\subsection{ The second family of codes $\mathcal{C}(m,2,L_2)$ when $m$ is singly-even }

We now consider $m$ is singly-even, which implies $R_m^*\cong   \F_{2^m}^*\bigoplus\F_{4^m}^*\bigoplus\F_{4^m}^*.$
A simple calculation shows that \begin{eqnarray*}
                                  R_m^* &=&\Bigg{\{}\sum_{i=0}^4h_iv^i: \sum_{i=0}^4h_i\neq 0,
(\omega^2h_1+\omega h_2+\omega h_3+\omega^2 h_4,\omega^2h_0+\omega h_1+\omega h_2+\omega^2 h_4)\neq(0,0), \\
                                   && (\omega h_1+\omega^2 h_2+\omega^2 h_3+\omega h_4,\omega h_0+\omega^2 h_1+\omega^2 h_2+\omega h_4)\neq(0,0)
\Bigg{\}}.
                                \end{eqnarray*}
For convenience, we adopt the following notations unless otherwise stated
in this section. Set $H=\sum_{i=0}^4h_i,H_1=\omega^2h_1+\omega h_2+\omega h_3+\omega^2 h_4,H_2=\omega^2h_0+\omega h_1+\omega h_2+\omega^2 h_4,H_3=\omega h_1+\omega^2 h_2+\omega^2 h_3+\omega h_4,H_4=\omega h_0+\omega^2 h_1+\omega^2 h_2+\omega h_4$.
Then $R_m^*=\{\sum_{i=0}^4h_iv^i: H\neq 0, (H_1,H_2)\neq(0,0),(H_3,H_4)\neq(0,0),H,H_1,H_2,H_3,H_4\in \F_{2^m} \}.$

 We are now ready to discuss the Lee weight of the
codewords of the abelian codes introduced above.

\noindent{\bf Theorem 4.4} Let $m$ be singly-even, then the set $\mathcal{C}(m,2,L_2)$ is a five-Lee-weight linear code of length $L_2=(2^m-1)(2^{2m}-1)^2$ and its weight distribution is given in Table II.
\begin{center}$\mathrm{Table~ II. }~~~\mathrm{weight~ distribution~ of}~ \mathcal{C}(m,2,L_2) $\\
\begin{tabular}{cccc||cc}
\hline
  Weight&&   & & Frequency  \\
  \hline
 0        & &   & & 1\\
  $5\times(2^{2m}-1)(2^{3m-1}-2^{2m-1}-2^{m-1})$        & &   &              &$2\times(2^m-1)(2^{2m}-1)$\\
   $5\times(2^m-1)(2^{2m-1}-1)2^{2m}$        & &   &              &$(2^{2m}-1)^2$\\
      $\frac{5\times[(2^m-1)(2^{2m}-1)^2+1]}{2}$        & &   &              &$(2^m-1)(2^{2m}-1)^2$\\
     $5\times(2^m-1)(2^{2m}-1)2^{2m-1}$  &    & &       &$2\times(2^{2m}-1)$ \\
   $5\times(2^{2m}-1)^22^{m-1}$  &    & &       &$2^m-1$ \\
  \hline
\end{tabular}
\end{center}

\begin{proof}
Set $h=h_0+h_1v+h_2v^2+h_3v^3+h_4v^4\in R_m^*$, then $H\neq0,(H_1,H_2)\neq(0,0),(H_3,H_4)\neq(0,0)$.
For $b=b_0+b_1v+b_2v^2+b_3v^3+b_4v^4\in R_m$, we have
\begin{eqnarray*}
  bh &=&[(b_0+b_1+b_2+b_3+b_4)H+(b_0+b_2+\omega^2b_3+\omega^2b_4 )H_1+
(b_1+b_4+\omega^2b_2+\omega^2b_3   )H_2\\
&&+
(b_0+b_2+\omega b_3+\omega b_4)H_3+
(b_1+b_4+\omega b_2+\omega b_3)H_4]\\
&& +[(b_0+b_1+b_2+b_3+b_4)H+(b_1+b_3+\omega^2b_0+\omega^2b_4 )H_1+
(b_0+b_2+\omega^2b_3+\omega^2b_4   )H_2\\
&&+
(b_1+b_3+\omega b_0+\omega b_4)H_3+
(b_0+b_2+\omega b_3+\omega b_4)H_4]v \\
&&+ [(b_0+b_1+b_2+b_3+b_4)H+(b_2+b_4+\omega^2b_0+\omega^2b_1)H_1+
(b_1+b_3+\omega^2b_0+\omega^2b_4   )H_2\\
&&+
(b_2+b_4+\omega b_0+\omega b_1)H_3+
(b_1+b_3+\omega b_0+\omega b_4)H_4]v^2\\
&&+  [(b_0+b_1+b_2+b_3+b_4)H+(b_0+b_3+\omega^2b_1+\omega^2b_2 )H_1+
(b_2+b_4+\omega^2b_0+\omega^2b_1   )H_2\\
&&+
(b_0+b_3+\omega b_1+\omega b_2)H_3+
(b_2+b_4+\omega b_0+\omega b_1)H_4]v^3 \\
&&+  [(b_0+b_1+b_2+b_3+b_4)H+(b_1+b_4+\omega^2b_2+\omega^2b_3 )H_1+
(b_0+b_3+\omega^2b_1+\omega^2b_2   )H_2\\
&&+
(b_1+b_4+\omega b_2+\omega b_3)H_3+
(b_0+b_3+\omega b_1+\omega b_2)H_4]v^4\\
&=:& B_0+B_1v+B_2v^2+B_3v^3+B_4v^4.
\end{eqnarray*}
So $$\Phi(Tr_m(bh))=(tr_m(B_0), tr_m(B_1),tr_m(B_2), tr_m(B_3),tr_m(B_4)).$$
Taking character sums, yields
\begin{eqnarray*}
  \Theta(b) &=&\sum_{i=0}^{4} \sum_{H\in \F_{2^m}^*} \sum_{(H_1,H_2)\neq(0,0),H_1,H_2\in \F_{2^m}} \sum_{(H_3,H_4)\neq(0,0),H_3,H_4\in \F_{2^m}} (-1)^{tr_m(B_i)  }.
\end{eqnarray*}

1) If $b=b_0+b_1v+b_2v^2+b_3v^3+b_4v^4\in R_m^*$, then $ \sum_{i=0}^4b_i\neq 0$ and $
(\omega^2b_1+\omega b_2+\omega b_3+\omega^2 b_4,\omega^2b_0+\omega b_1+\omega b_2+\omega^2 b_4)\neq(0,0), (\omega b_1+\omega^2 b_2+\omega^2 b_3+\omega b_4,\omega b_0+\omega^2 b_1+\omega^2 b_2+\omega b_4)\neq(0,0)$.
What is more, we obtain $\Theta(b)=-5.$ By Equation (1), we get $$w_L(ev(b))=\frac{5\times[(2^m-1)(2^{2m}-1)^2+1]}{2}.$$

2) If $ \sum_{i=0}^4b_i=0$ and
$(\omega^2b_1+\omega b_2+\omega b_3+\omega^2 b_4,\omega^2b_0+\omega b_1+\omega b_2+\omega^2 b_4)\neq(0,0), (\omega b_1+\omega^2 b_2+\omega^2 b_3+\omega b_4,\omega b_0+\omega^2 b_1+\omega^2 b_2+\omega b_4)\neq(0,0)$,
 then $b=b_0+b_1v+b_2v^2+b_3v^3+b_4v^4$ is a zero-divisor. Furthermore, $\Theta(b)=5\times(2^m-1).$
By Equation (1), we get $$w_L(ev(b))=5\times(2^m-1)(2^{2m-1}-1)2^{2m}.$$

3) If $ \sum_{i=0}^4b_i\neq 0$ and $
(\omega^2b_1+\omega b_2+\omega b_3+\omega^2 b_4,\omega^2b_0+\omega b_1+\omega b_2+\omega^2 b_4)=(0,0), (\omega b_1+\omega^2 b_2+\omega^2 b_3+\omega b_4,\omega b_0+\omega^2 b_1+\omega^2 b_2+\omega b_4)\neq(0,0),$
 then $\Theta(b)=5\times(2^{2m}-1).$

  When $ \sum_{i=0}^4b_i\neq 0$ and $
(\omega^2b_1+\omega b_2+\omega b_3+\omega^2 b_4,\omega^2b_0+\omega b_1+\omega b_2+\omega^2 b_4)\neq(0,0), (\omega b_1+\omega^2 b_2+\omega^2 b_3+\omega b_4,\omega b_0+\omega^2 b_1+\omega^2 b_2+\omega b_4)=(0,0)$,
 then $\Theta(b)=5\times(2^{2m}-1).$
 By Equation (1), we get $$w_L(ev(b))=5\times(2^{2m}-1)(2^{3m-1}-2^{2m-1}-2^{m-1}).$$

4) If $ \sum_{i=0}^4b_i= 0$ and $
(\omega^2b_1+\omega b_2+\omega b_3+\omega^2 b_4,\omega^2b_0+\omega b_1+\omega b_2+\omega^2 b_4)=(0,0), (\omega b_1+\omega^2 b_2+\omega^2 b_3+\omega b_4,\omega b_0+\omega^2 b_1+\omega^2 b_2+\omega b_4)\neq(0,0),$
 then $\Theta(b)=-5\times(2^m-1)(2^{2m}-1).$

 When $ \sum_{i=0}^4b_i= 0$ and $
(\omega^2b_1+\omega b_2+\omega b_3+\omega^2 b_4,\omega^2b_0+\omega b_1+\omega b_2+\omega^2 b_4)\neq(0,0), (\omega b_1+\omega^2 b_2+\omega^2 b_3+\omega b_4,\omega b_0+\omega^2 b_1+\omega^2 b_2+\omega b_4)=(0,0)$,
 then $\Theta(b)=-5\times(2^m-1)(2^{2m}-1).$
 By Equation (1), we get $$w_L(ev(b))=5\times(2^m-1)(2^{2m}-1)2^{2m-1}.$$

5) If $b=b_0+b_1v+b_2v^2+b_3v^3+b_4v^4$ is a zero-divisor, where $ \sum_{i=0}^4b_i\neq 0$ and $
(\omega^2b_1+\omega b_2+\omega b_3+\omega^2 b_4,\omega^2b_0+\omega b_1+\omega b_2+\omega^2 b_4)=(0,0), (\omega b_1+\omega^2 b_2+\omega^2 b_3+\omega b_4,\omega b_0+\omega^2 b_1+\omega^2 b_2+\omega b_4=(0,0),$
 then $b=\beta(1+v+v^2+v^3+v^4)$, where $\beta\in \F_{2^m}^*$, and $ bh =\beta H(1+v+v^2+v^3+v^4)$.
So $\Phi(Tr_m(bh))=(tr_m(\beta H), tr_m(\beta H),tr_m(\beta H), tr_m(\beta H),tr_m(\beta H))$.
Taking  character sums, yields
\begin{eqnarray*}
  \Theta(b) &=&5\sum_{H\in \F_{2^m}^*} \sum_{(H_1,H_2)\neq(0,0),H_1,H_2\in \F_{2^m}} \sum_{(H_3,H_4)\neq(0,0),H_3,H_4\in \F_{2^m}} (-1)^{tr_m(\beta H)  }=-5\times(2^{2m}-1)^2.
\end{eqnarray*}By Equation (1), we get $$w_L(ev(b))=5\times(2^{2m}-1)^22^{m-1}.$$

Note that $ \sum_{i=0}^4b_i= 0$ and $
(\omega^2b_1+\omega b_2+\omega b_3+\omega^2 b_4,\omega^2b_0+\omega b_1+\omega b_2+\omega^2 b_4)= (\omega b_1+\omega^2 b_2+\omega^2 b_3+\omega b_4,\omega b_0+\omega^2 b_1+\omega^2 b_2+\omega b_4)=(0,0),$
 i.e.,
$b = 0,$ corresponds to the zero codewords, and then its Lee weight is 0.
\end{proof}\vspace*{0.2cm}

From Theorem 4.4, we have constructed a binary linear code of length $s=5\times(2^m-1)(2^{2m}-1)^2$, of dimension $5m$, with five nonzero weights $w_1<w_2<w_3<w_4<w_5$ of values
$$w_1=5\times(2^{2m}-1)(2^{3m-1}-2^{2m-1}-2^{m-1}), ~~~w_2=5\times(2^m-1)(2^{2m-1}-1)2^{2m},~~~$$
$$w_3=\frac{5\times[(2^m-1)(2^{2m}-1)^2+1]}{2},~~~w_4=5\times(2^m-1)(2^{2m}-1)2^{2m-1},~
~~w_5=5\times(2^{2m}-1)^22^{m-1},$$
with respective frequencies
$$f_1=2\times(2^m-1)(2^{2m}-1),~~f_2=(2^{2m}-1)^2,~~f_3=(2^m-1)(2^{2m}-1)^2,~~f_4=2\times(2^{2m}-1),~~f_5=2^m-1.$$

\subsection{ The third family of codes $\mathcal{C}(m,2,L_3)$ when $m$ is doubly-even }

Writing $\eta_0=1+v+v^2+v^3+v^4,~ \eta_1=1+\epsilon^4v+\epsilon^3v^2+\epsilon^2v^3+\epsilon v^4,
 ~ \eta_2=1+\epsilon^3v+\epsilon v^2+\epsilon^4v^3+\epsilon^2 v^4, ~\eta_3=1+\epsilon^2v+\epsilon^4 v^2+\epsilon v^3+\epsilon^3 v^4$ and $\eta_4=1+\epsilon v+\epsilon^2 v^2+\epsilon^3 v^3+\epsilon^4 v^4,$ where $\epsilon^5=1$. According to Chinese Remainder Theorem, we then obtain $R_m=\eta_0\F_{2^m}\oplus\eta_1\F_{2^m}\oplus\eta_2\F_{2^m}\oplus\eta_3\F_{2^m}\oplus\eta_4\F_{2^m}$ in the case of $m$ is doubly-even. Hence, we have $R_m^*\cong\F_{2^m}^*\bigoplus \F_{2^m}^*\bigoplus\F_{2^m}^*\bigoplus\F_{2^m}^*\bigoplus\F_{2^m}^*$.
 It means that $R_m^*=\{\eta_0r_0+ \eta_1r_1+\eta_2r_2+\eta_3r_3+\eta_4r_4: r_j\in \F_{2^m}^*, j=0,1,2,3,4 \}$. Now, we investigate the Lee weight of codewords of $\mathcal{C}(m,2,L_3)$ in this case.\\

 \noindent{\bf Theorem 4.5} Let $m$ be doubly-even, then the set $\mathcal{C}(m,2,L_3)$ is a five-Lee-weight linear code of length $L_3=(2^m-1)^5$ and its weight distribution is given in Table III.
\begin{center}$\mathrm{Table~ III. }~~~\mathrm{weight~ distribution~ of}~ \mathcal{C}(m,2,L_3) $\\
\begin{tabular}{cccc||cc}
\hline
  Weight&&   & & Frequency  \\
  \hline

  0        & &   & & 1\\
 $5\times(2^m-1)^3(2^m-2)2^{m-1} $  &    & &       &$10\times(2^m-1)^2$ \\
  $5\times(2^m-1)(2^m-2)(2^{2m}-2^{m+1}+2)2^{m-1}$        & &   &              &$5\times(2^m-1)^4$\\
  $\frac{ 5\times[(2^m-1)^5+1]}{2} $        & &   &              &$(2^m-1)^5$\\
   $ 5\times(2^m-1)^2(2^{3m-1}-2^{2m-1}+2^{m-1}) $  &    & &      &$10\times(2^m-1)^3$\\
  $ 5\times(2^m-1)^42^{m-1} $  &    & &   &$5\times(2^m-1)$\\
  \hline
\end{tabular}
\end{center}

\begin{proof}
Set $r=\eta_0r_0+ \eta_1r_1+\eta_2r_2+\eta_3r_3+\eta_4r_4\in R_m^*$, where $r_j\in \F_{2^m}^*, j=0,1,2,3,4$.
For $a=\eta_0 a_0+\eta_1a_1+\eta_2a_2+\eta_3a_3+\eta_4a_4\in R_m$, where $ a_j\in \F_{2^m},j=0,1,2,3,4$, then
$ar=\eta_0a_0r_0+ \eta_1a_1r_1+\eta_2a_2r_2+\eta_3a_3r_3+\eta_4a_4r_4$ and
\begin{eqnarray*}
 \Phi( Tr_m(ar)) &=& (tr(a_0r_0+a_1r_1+a_2r_2+a_3r_3+a_4r_4),
tr(a_0r_0+\epsilon^4 a_1r_1+\epsilon^3a_2r_2+\epsilon^2a_3r_3+\epsilon a_4r_4), \\
&& tr(a_0r_0+\epsilon^3 a_1r_1+\epsilon a_2r_2+\epsilon^4 a_3r_3+\epsilon^2a_4r_4),
  tr(a_0r_0+\epsilon^2 a_1r_1+\epsilon^4a_2r_2+\epsilon a_3r_3+\epsilon^3a_4r_4),\\
&&tr(a_0r_0+\epsilon a_1r_1+\epsilon^2a_2r_2+\epsilon^3a_3r_3+\epsilon^4a_4r_4))\\
&=:& (A'_0,A'_1,A'_2,A'_3,A'_4).
\end{eqnarray*}
Taking character sums, yields
$$\Theta(a)=\sum_{i=0}^4\sum_{r_0,r_1,r_2\in \F_{2^m}^*}\sum_{r_3,r_4\in \F_{2^m}^*}(-1)^{tr_m(A'_i)}.$$

If $a=\eta_0 a_0+\eta_1a_1+\eta_2a_2+\eta_3a_3+\eta_4a_4\in R_m^*$, where $ a_j\in \F_{2^m}^*,j=0,1,2,3,4$, then $\Theta(a)=-5$. So $w_L(ev(a))=\frac{ 5\times[(2^m-1)^5+1]}{2} .$

 Next, we consider $a\in R_m\backslash\{R_m^*\}$. It is easily check that $a=0$ corresponds to the zero codewords, and then its Lee weight is 0.
If $a=\eta_ia_i, i\in \{0,1,2,3,4\}$, then $\Theta(a)=5\times(2^m-1)$.
Further, $w_L(ev(a))=5\times(2^m-1)(2^m-2)(2^{2m}-2^{m+1}+2)2^{m-1}$ by Equation (1). If $a=\eta_ia_i+\eta_ja_j, i,j\in \{0,1,2,3,4\},i\neq j$, then $\Theta(a)=-5\times(2^m-1)^2$. By Equation (1), we have  $w_L(ev(a))= 5\times(2^m-1)^2(2^{3m-1}-2^{2m-1}+2^{m-1})$.
If $a=\eta_ia_i+\eta_ja_j+\eta_ka_k,i,j,k\in \{0,1,2,3,4\},$ where $i,j,k$ are pairwise different, then $\Theta(a)=5\times(2^m-1)^3$. What is more, $w_L(ev(a))= 5\times(2^m-1)^3(2^m-2)2^{m-1}$. If $a=\eta_0 a_0+\eta_1a_1+\eta_2a_2+\eta_3a_3+\eta_4a_4$, where $a_j=0,a_i\in \F_{2^m}^*,j\in \{0,1,2,3,4\}, i\in\{0,1,2,3,4\}\backslash\{j\}$, then we can obtain $\Theta(a)=-5\times(2^m-1)^4.$
By Equation (1), we can easily get $w_L(ev(a))=5\times(2^m-1)^42^{m-1}$.
\end{proof}\vspace*{0.2cm}

By Theorem 4.5, we have constructed a binary linear code of length $s=5\times(2^m-1)^5$, of dimension $5m$, with five nonzero weights $w_1<w_2<w_3<w_4<w_5$ of values
$$w_1=5\times(2^m-1)^3(2^m-2)2^{m-1} , ~~w_2=5\times(2^m-1)(2^m-2)(2^{2m}-2^{m+1}+2)2^{m-1},~~w_3=\frac{ 5\times[(2^m-1)^5+1]}{2}$$
$$w_4=5\times(2^m-1)^2(2^{3m-1}-2^{2m-1}+2^{m-1}),~w_5=5\times(2^m-1)^42^{m-1} ,$$
with respective frequencies
$$f_1=10\times(2^m-1)^2,~~f_2=5\times(2^m-1)^4,~~f_3=(2^m-1)^5,~~f_4=10\times(2^m-1)^3,~~f_5=5\times(2^m-1).$$


Note that $L=L_1=L_2=L_3=|R^*_m|.$ Next, we study their optimality.\\
\noindent{\bf Theorem 4.6} The three-weight binary linear code $\Phi(\mathcal{C}(m,2,L_1))$, for $m>6$ and $m$ is odd, is optimal.
\begin{proof}
Recall the Griesmer bound \cite{G}. If $[N,K,d]$ are the parameters of a linear binary code, then $$\sum_{j=0}^{K-1}\Big\lceil\frac{d}{2^j}\Big\rceil\leq N.$$
In our situation $N=5\times(2^{5m}-2^{4m}-2^{m}+1),K =5m,d =5\times(2^{5m-1}-2^{4m-1}-2^{m-1}).$
The ceiling
function takes the following values depending on the position of $j:$
\begin{itemize}
 \item $0\le j\le m-1 \Rightarrow \lceil \frac{d+1}{2^j} \rceil =5\times(2^{4m}-2^{3m}-1)2^{m-1-i}+1,$
 \item $j= m \Rightarrow \lceil \frac{d+1}{2^j} \rceil =5\times(2^{4m-1}-2^{3m-1})-2,$
  \item $j= m+1 \Rightarrow \lceil \frac{d+1}{2^j} \rceil =5\times(2^{4m-2}-2^{3m-2})-1,$
 \item $m+2\le j\le 4m-1 \Rightarrow \lceil \frac{d+1}{2^j} \rceil =5\times(2^{m}-1)2^{4m-1-i},$

  \item $j= 4m \Rightarrow \lceil \frac{d+1}{2^j} \rceil =5\times2^{m-1}-2,$
  \item $j= 4m+1 \Rightarrow \lceil \frac{d+1}{2^j} \rceil =5\times2^{m-2}-1,$
 \item $4m+2\le j\le 5m-1  \Rightarrow \lceil \frac{d+1}{2^j} \rceil =5\times2^{5m-1-i}.$
\end{itemize}
Thus,
\begin{eqnarray*}
  \sum_{j=0}^{K-1}\Big\lceil\frac{d+1}{2^j}\Big\rceil &=&   \sum_{j=0}^{5m-1} 5\times2^{5m-1-i}
 -\sum_{j=0}^{4m-1}5\times2^{4m-1-i}   -\sum_{j=0}^{m-1}5\times2^{m-1-i}+m -6\\
   &=&5\times(2^{5m}-1)- 5\times(2^{4m}-1)- 5\times(2^{m}-1)+m- 6\\
   &=& 5\times2^{5m}- 5\times2^{4m}- 5\times2^{m} +m- 1.
\end{eqnarray*}
By simply calculation, we have $5\times2^{5m}- 5\times2^{4m}- 5\times2^{m} +m- 1-N> 0 $ when $m>6$.
This proof is completed.
\end{proof}\vspace*{0.2cm}
\noindent{\bf Example 4.7}
 Let $m=2$. By Theorem 4.4, we obtain a five-weight binary linear code of length 1215, of dimension 10, with five nonzero weights 1650, 1680, 1690, 1800, 2250, and frequencies 90, 225, 675, 30 and 3, respectively.

\noindent{\bf Example 4.8}  Taking $m=3$. Then we get a three-weight binary linear code of dimension 15, with three nonzero weights 71660, 71680, 81900, and frequencies 28665, 4095 and 7, respectively.

%
%

\section{ The dual code}
We compute the dual distance of $\mathcal{C}(m,2,|R_m^*|).$ A property of the trace that we need is that it is nondegenerate. The proof of the following lemma is similar to that in \cite{SLS2}, so we omit it here.

\noindent{\bf Lemma 5.1}
 If for all $a \in R_m,$ we have that $Tr_m(ax)=0,$ then $x=0.$

Combining the sphere-packing bound and Lemma 5.1, we can get the following results.

\noindent\textbf{Theorem 5.2}
The dual Lee distance $d'$ of $\mathcal{C}(m,2,|R_m^*|)$ is $2.$

\begin{proof}
 First, we show that $d'<3.$ If not, we can apply the sphere-packing bound to $\Phi(\mathcal{C}(m,2,|R_m^*|)^\bot),$ to obtain $$2^{5m}\ge 1+s.$$
 \begin{itemize}
 \item $ m$ is odd, for $\Phi(\mathcal{C}(m,2,L_1)^\bot) $, then $$2^{5m}\ge 1+s=1+5\times(2^m-1)(2^{4m}-1),$$ or, after expansion
 $$5\times(2^{4m}+2^{m})\ge 6+2^{5m+2}. $$
 Dropping the $6$ in the RHS, and dividing both sides by $2^{m},$ we find that this inequality would imply $5\times(2^{3m}+1)>2^{5m+2},$ i.e., $5>2^{3m}(2^{2m+2}-5)$. Contradiction with $m\geq1$, which implies $d'<3.$
  \item $m$ is singly-even, for $\Phi(\mathcal{C}(m,2,L_2)^\bot) $, then
  \begin{equation}\label{2}
    2^{5m}\ge 1+s=1+5\times(2^m-1)(2^{2m}-1)^2=1+5\times(2^{5m}-2^{4m}-2^{3m+1}+2^{2m+1}+2^m-1).
  \end{equation}
  When $m=2$, then $2^{5m}<1+5\times(2^{5m}-2^{4m}-2^{3m+1}+2^{2m+1}+2^m-1).$ Now, we consider $m\geq6$.
 From Equation (2), we have
   \begin{eqnarray*}
     0 &\geq& 2^{5m+2}-5\times(2^{4m}+2^{3m+1} )+5\times(2^{2m+1}+2^m)-4 \\
      &\geq&2^{3m+1}(2^{2m+1}-5\times2^{m-1}-5 ).
   \end{eqnarray*}
 Contradiction with $2^{2m+1}-5\times2^{m-1}-5 =2^{m-1}(2^{m+2}-5)-5>0$ when $m\geq6.$
 So $d'<3.$

 \item $m$ is doubly-even, for $\Phi(\mathcal{C}(m,2,L_3)^\bot) $, then
 \begin{eqnarray*}
   2^{5m} &\ge& 1+5\times(2^m-1)^5\\
  & =&5\times(2^{5m}-2^{4m+2}-2^{4m}+2^{3m+3}+2^{3m+1}-2^{2m+3}-2^{2m+1}+2^{m+2}+2^m)-4 \\
    &>& 5\times(2^{5m}-2^{4m+2}-2^{4m}).
 \end{eqnarray*}
\end{itemize}
Note that $5\times(2^{5m}-2^{4m+2}-2^{4m})- 2^{5m}=2^{4m}(2^{m+2}-25)>0$ for $m\geq4.$
It is contradiction, which implies $d'<3.$

 Next, we check that $d'= 2,$ by showing that $\mathcal{C}(m,2,|R_m^*|)^\bot,$ does not contain a word of Lee weight one.
 If it does, let us assume first that it has value $\gamma v^j$ at some $x \in R_m^*,$ where $\gamma\in \F_{2^m}^*,j\in \{0,1,2,3,4\}.$
 This implies that $\forall a \in R_m,\gamma v^j
 Tr_m(ax)=0,$ then we must have $Tr_m(\gamma ax)=0$ since $Tr_m()$ is a linear map, and by using Lemma 5.1, we know $x=0.$ Contradiction with $x\in R_m^*.$ So $d'=2.$
\end{proof}


\section{Application to secret sharing schemes}
Secret sharing is an important topic of cryptography, which has been studied for over thirty years. In this section, we will study the secret sharing schemes based on linear codes studied in this paper.

Originally secret sharing was motivated
by the problem of sharing a secret digital key. In order to keep the secret efficiently and
safely, Shamir and Blakley introduced secret sharing schemes (SSS) in 1979. An SSS based on error-correcting codes was introduced by Massey, and the minimal coalitions
in such a scheme were characterized as a function of the minimal vectors, to be defined next, of the dual code.
   A \emph{minimal codeword} of a linear code $C$ is a nonzero codeword that does not cover any other nonzero codeword. Recall
   that a vector $x$ {\em covers} a vector $y$ if $s(x)$ contains $s(y)$, where $s(y)$, the support $s(y)$ of a vector $y\in \F_p^s$, is defined as the set of indices where it is nonzero.
   Although the minimal codewords of a given linear code is hard to determine in general, there is a numerical condition derived in \cite{AB}, bearing on the weights of the code,
   that is easy to check, once the weight distribution is known. By recalling the following result of Ashikhmin and Barg (see \cite{AB}), it is shown that all
nonzero codewords of linear code are minimal.

\noindent{\bf Lemma 6.1 }(Ashikhmin-Barg) Given a $p$-ary code $C,$  denote by $w_0$ and $w_{\infty}$ the minimum and maximum nonzero weights of $C$, respectively. If
$$\frac{w_0}{w_{\infty}}>\frac{p-1}{p},$$ then every nonzero codeword of $C$ is minimal. \


\noindent{\bf Theorem 6.2} Let $m$ be odd,
 then all the nonzero codewords of $\Phi(\mathcal{C}(m,2,L_1))$, for $m>1$, are minimal.
\begin{proof}
 By the preceding Lemma 6.1 with $w_0= 5\times(2^{5m-1}-2^{4m-1}-2^{m-1}),$ and $w_{\infty}=5\times(2^{5m-1}-2^{m-1}).$ Rewriting the inequality of Lemma 6.1 as $2\omega_1>\omega_2$, and dividing both sides by $5\times 2^{m-1}$, we obtain
 $$2^{4m}>2^{3m+1}+1.$$
 The condition follows from the fact that $2^{m-1}>1$ when $m>1.$
 Hence the result follows.
\end{proof}

\noindent{\bf Theorem 6.3} Let $m$ be singly-even,
 then all the nonzero codewords of $\Phi(\mathcal{C}(m,2,L_2))$, for $m>1$, are minimal.
\begin{proof}
 We use Lemma 6.1 with $w_0= 5\times(2^{2m}-1)(2^{3m-1}-2^{2m-1}-2^{m-1})$ and $w_{\infty}=5\times(2^{2m}-1)^22^{m-1}.$ Rewriting the inequality of Lemma 6.1 as $2w_0>w_\infty,$ and dividing both sides by $5\times(2^m-1) 2^{m-1}$, we obtain
$$2\times(2^{2m}-1)>2^{2m}-2^m-1,$$
namely, $2^{2m}+2^m>1$. Note that $2^{2m}+2^m>1$ for a positive integer $m.$ The result follows.
\end{proof}

\noindent{\bf Theorem 6.4} Let $m$ be doubly-even,
 then all the nonzero codewords of $\Phi(\mathcal{C}(m,2,L_3))$, for $m>1$, are minimal.
\begin{proof}
 We use Lemma 6.1 with $w_0= 5\times(2^{m}-1)^3(2^{m}-2)2^{m-1},$ and $w_{\infty}=5\times(2^{m}-1)^42^{m-1}.$ Rewriting the inequality of Lemma 6.1 as $2w_0>w_\infty,$ and dividing both sides by $5\times(2^m-1)^3 2^{m-1}$, we obtain
$$2\times(2^m-2)>2^m-1,$$
namely, $2^m>3$. Noting that $2^{m}>3$ when $m>1.$ So the result follows.
\end{proof}

 The Massey's scheme is a construction of an SSS using a code $C$ of length $s$ over $\F_p.$  In essence,
the secret is carried by the first coordinate of a codeword,
and the coalitions correspond to supports of codewords in
the dual code with a one in that coordinate. It is worth mentioning that in some special cases, that is, when all nonzero codewords are minimal, it was shown in \cite{DY2} that there is the following alternative, depending on $d'$:
\begin{itemize}
 \item If $d'\ge 3,$ then the SSS is \emph{``democratic''}: every user belongs to the same number of coalitions,
 \item If $d'=2,$  then the SSS is  \emph{``dictatorial''}: some users belong to every coalition.
\end{itemize}
Depending on the application, one or the other situation might be more suitable.
By Theorems 4.3, 4.4,  4.5 and  5.2, we see that for some values of the parameters, a SSS built on $\Phi(\mathcal{C}(m,p,|R_m^*|))$ is dictatorial.
\section{Conclusion}
Trace codes over fields are a well-known source of constructions for few weights codes. In the present work, we have extended the notion of trace codes from fields to rings. On the basis of the linear Gray map we defined, we constructed a family of three-weight binary linear codes, which are optimal by using the Griesmer bound, and two families of five-weight binary
linear codes. These codes are abelian, and quasi-cyclic, but not visibly cyclic. Finally, an application to secret sharing schemes is given.
It is worth exploring more general constructions by varying the alphabet of the code, or the defining set of the trace code.
Compared with linear codes in \cite{DD,DLLZ,DY,HY1,ZDL}, the codes in this paper have different weight distribution.


\section*{References}

\bibliography{mybibfile}

\end{document}